\documentclass[english,aps,superscriptaddress,preprintnumbers,reprint,footinbib,amsmath,amssymb,prbr]{revtex4-2}
\usepackage[latin9]{inputenc}
\usepackage{amsmath}
\usepackage{amssymb}
\usepackage{graphicx}
\usepackage{esint}
\usepackage{xcolor}

\makeatletter

\usepackage{textcomp}
\usepackage[toc]{appendix}

\@ifundefined{textcolor}{}{%
 \definecolor{BLACK}{gray}{0}
 \definecolor{WHITE}{gray}{1}
 \definecolor{RED}{rgb}{1,0,0}
 \definecolor{GREEN}{rgb}{0,1,0}
 \definecolor{BLUE}{rgb}{0,0,1}
 \definecolor{CYAN}{cmyk}{1,0,0,0}
 \definecolor{MAGENTA}{cmyk}{0,1,0,0}
 \definecolor{YELLOW}{cmyk}{0,0,1,0}
}

\usepackage{ulem}

\usepackage{aecompl}

\usepackage{epsfig}\usepackage{dcolumn}\usepackage{bm}

\usepackage{babel}

\makeatother

\usepackage{babel}
\begin{document}
\title{{Atomic frustration-based twistronics}}
\author{W. N. Mizobata}
\affiliation{São Paulo State University (Unesp), School of Engineering, Department
of Physics and Chemistry, 15385-000, Ilha Solteira-SP, Brazil}
\author{J. E. Sanches}
\affiliation{São Paulo State University (Unesp), School of Engineering, Department
of Physics and Chemistry, 15385-000, Ilha Solteira-SP, Brazil}
\author{M. Penha}
\affiliation{São Paulo State University (Unesp), School of Engineering, Department
of Physics and Chemistry, 15385-000, Ilha Solteira-SP, Brazil}
\author{W. C. Silva}
\affiliation{São Paulo State University (Unesp), School of Engineering, Department
of Physics and Chemistry, 15385-000, Ilha Solteira-SP, Brazil}
\author{\\
 C. A. Carvalho}
\affiliation{São Paulo State University (Unesp), School of Engineering, Department
of Physics and Chemistry, 15385-000, Ilha Solteira-SP, Brazil}
\author{M. S. Figueira}
\affiliation{Instituto de Física, Universidade Federal Fluminense, 24210-340, Niterói,
Rio de Janeiro, Brazil}
\author{M. de Souza}
\affiliation{São Paulo State University (Unesp), IGCE, Department of Physics, 13506-970,
Rio Claro-SP, Brazil}
\author{A. C. Seridonio}
\email[corresponding author: ]{antonio.seridonio@unesp.br}

\affiliation{São Paulo State University (Unesp), School of Engineering, Department
of Physics and Chemistry, 15385-000, Ilha Solteira-SP, Brazil}
\affiliation{São Paulo State University (Unesp), IGCE, Department of Physics, 13506-970,
Rio Claro-SP, Brazil}
\begin{abstract}
We theoretically investigate atomic frustrated states in diatomic
molecules hosted by the bilayer graphene setup twisted by the first
magic angle and with broken inversion symmetry in the Dirac cones
of the system mini Brillouin zones. Such states show local spectral
features typically from uncoupled atoms, but counterintuitively, they
also exhibit nonlocal molecular correlations, which turn them into
atomically frustrated. By considering a particle-hole symmetric molecule
in the Moiré superlattice length-scale, we reveal distinctly from
the metallic Weyl counterparts, a molecular zero mode atomically frustrated
at the spectral densities of the dimer's atoms. To this end, a strong
metallic phase with a plateau in the density of states established
by the broken inversion symmetry, together with pronounced blue and
red shifts in the molecular levels, due to the magic angle condition,
should occur synergistically with atomic Coulomb correlations. Consequently,
an entire collapse of these molecular peaks into a single one atomically
frustrated, taking place exactly at the Fermi energy, becomes feasible
just by tuning properly opposite gate voltages attached to the graphene
monolayers. Therefore, we propose that unusual molecular bindings
can be engineered via the twistronics of the bilayer graphene system,
in particular, if its metallic phase is fully established.
\end{abstract}
\maketitle

\section{Introduction}

In the twisted bilayer graphene (TBG) system, astonishing experimental
findings have revealed delicate interplay between the strong correlated
insulating and superconducting states\cite{Herrero1,Herrero2}. Consequently,
this scenario has stimulated substantial theoretical focus on such a matter\cite{T1,T2,T3,T4,T5,T6,T7,T8,T9,T10,T11,T12}.
Essentially, the TBG setup consists of two stacked graphene monolayers
misaligned to each other by a twist angle $\theta$\cite{EXP1,EXP2,EXP3},
from which the band-structure becomes highly dependent. Particular
attention should be paid to slight twists, once they yield the Moiré
superlattice represented by an effective honeycomb pattern, which
is a result from the interference between those found in the graphene
monolayers. Amazingly, twist angles are then capable of changing dramatically
the underlying Dirac Physics. This is noticed when the Fermi velocity
vanishes, in particular, by tuning the twist angles to the so-called
magic angles, which lead to flat bands in the domain of low energies,
namely, at the corners of the mini Brillouin zones\cite{TT1,TT2,TT3,CastroNetoPRB,TT5,TT6,TT7,TT8,TT9,TT10,Review,CastroNetoPRL,EJMele}.
As the TBG system is naturally a platform with a huge number of degrees
of freedom, due to an expressive quantity of carbon atoms per unit
cell and a long-range lattice parameter for the Moiré superlattice,
the determination of the TBG electronic properties constitutes a hard
task to perform. In this context, two common theoretical strategies
are usually employed to overcome such an issue\cite{CastroNetoPRB,CastroNetoPRL,EJMele}.
The first alternative is based on the continuum model\cite{CastroNetoPRB,CastroNetoPRL},
which individually accounts for the Hamiltonians from the misaligned graphene monolayers and that for their interlayer couplings.
However, such an approach still exhibits a significant number of energy bands. Unfortunately as a result,
the continuum model does not allow a simple description of the system in case of considering impurity atoms.
Interestingly enough, as the flat bands are within the low energy
domain, excited bands can be safely neglected, if we are interested
in the Dirac Physics. As aftermath, Wannier wave functions centered
around the AB and BA stacking regions of the Moiré superlattice can
be assumed, as a second theoretical strategy\cite{EJMele}. By means of such a procedure, a hexagonal
effective lattice emerges to better handle impurity atoms, once the emulation of the TBG is performed
by an effective graphene monolayer in the Moiré length-scale. It is worth mentioning that, differently from the continuum model where the pieces of the TBG system are treated separately, in such a superlattice description, the Dirac cones at the corners of the mini Brillouin zones, then wrap up all these parts entirely into a renormalized Fermi velocity, depending upon the twist angle and interlayer couplings\cite{CastroNetoPRB,Review,CastroNetoPRL,EJMele}.

\begin{figure*}[ht]
\centering\includegraphics[width=0.78\textwidth,height=0.44\textheight]{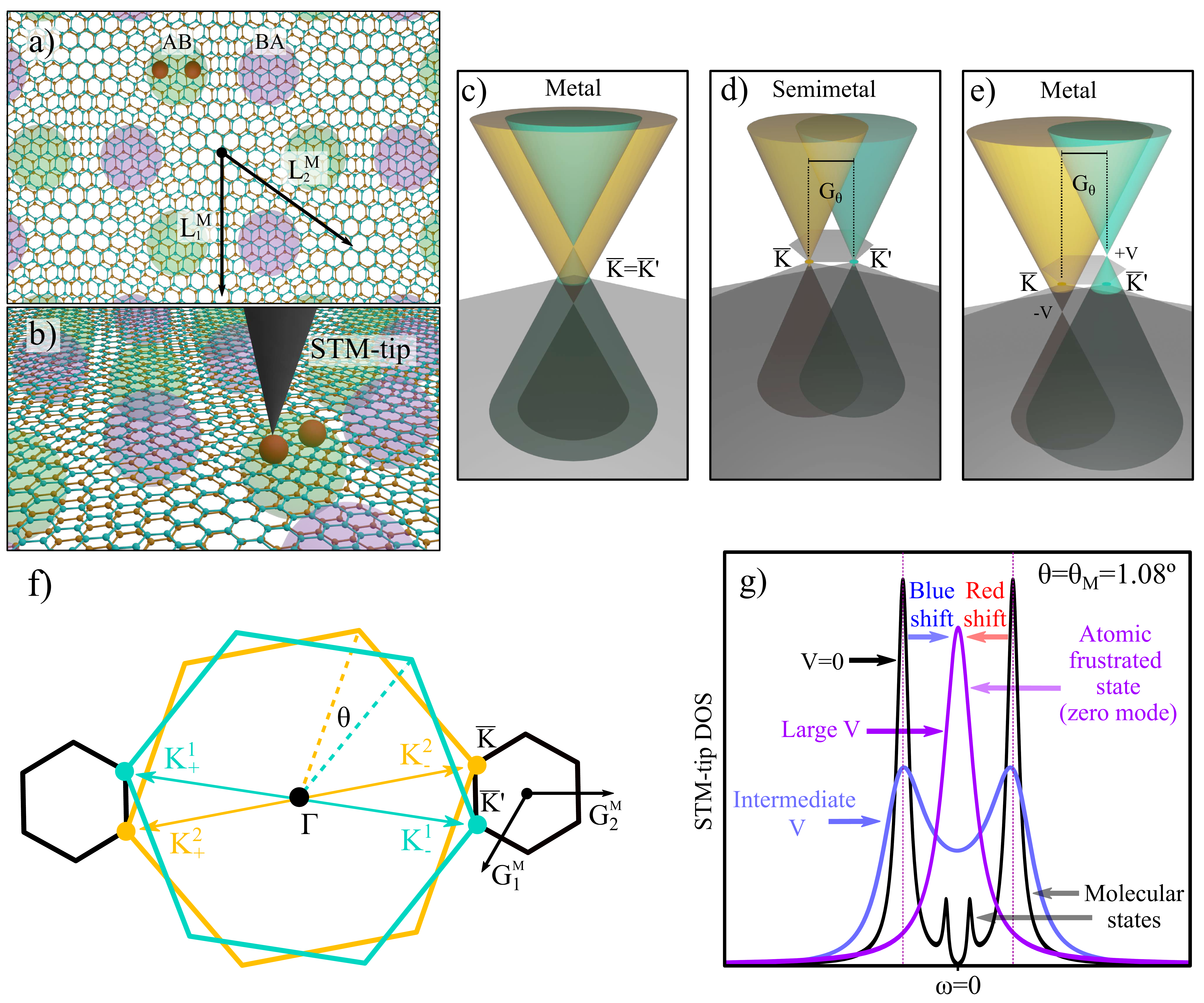}\caption{\label{fig:Pic1} {(Color online) Panels a) and f) were prepared after Ref.\cite{EJMele}. Panel (a): Top view for the TBG
system and its Moiré superlattice hexagonal pattern for small twist
angles $\theta.$ The base vectors are {$\textbf{L}_{1}^{\text{M}}$}
and {$\textbf{L}_{2}^{\text{M}}$}. The AB (green) and BA (purple) stacking regions
represent the Moiré superatoms and the diatomic molecule, sketched by the pair of red spheres, couples to the AB site. Panel (b): Lateral perspective of panel
(a) with an STM-tip above one atom of the dimer. The tip, via the
differential conductance, measures the density of states (DOS) of such an atom. Panel
(c): Band-structure for the untwisted system for the AA stacking,
with coincident Dirac cones at the valley corners {$\overline{\textbf{K}}=\overline{\textbf{K}'}$}
arising from the top and bottom Brillouin zones of the monolayers, respectively. These Dirac points appear shifted in energy, due to the broken inversion
symmetry in the metallic phase\cite{Review}, in analogy to the metallic
Weyl counterparts\cite{Frustration}. Panel (d): The twist restores
the semimetallic character of the system, but now splits the Dirac
points with {$\overline{\textbf{K}}\protect\neq\overline{\textbf{K}'}$}
in momentum space by $G_{\theta}=\frac{8\pi\text{sin}\frac{\theta}{2}}{3a}$,
where $a$ is the lattice parameter of the graphene monolayer. Panel
(e): External gate voltages attached to the top ($+V$) and bottom
($-V$) monolayers induce the breaking down of the inversion symmetry
with metallic character. Panel (f): Mini Brillouin zones with reciprocal
base vectors {$\textbf{G}_{1}^{\text{M}}$} and {$\textbf{G}_{2}^{\text{M}}$},
arising from the twist between the top ($1$) and bottom ($2$) Brillouin
zones of the graphene monolayers, which have as valley corners {$\textbf{K}_{+}^{1},\textbf{K}_{-}^{1},\textbf{K}_{+}^{2}$}
and {$\textbf{K}_{-}^{2}$}. Panel (g): Key finding of this work:
for the first twist magic angle $\theta=\theta_{M}=1.08^{{^{\circ}}},$
a crossover in the particle-hole symmetric profile in the spectral
DOS for the dimer's atoms is observed. It evolves
from a pattern with ordinary molecular states (black lineshape) towards
that characterized by a molecular single zero mode ($\omega=0$) (purple
lineshape), which is atomically frustrated {[}see Fig.\ref{fig:Pic3}{]}.
This zero mode is the aftermath of the merge between the ordinary
molecular states, which is due to the blue and red shifts enhancement
of their resonant peak positions, in particular, upon increasing the
breaking down of the inversion symmetry of the Dirac cones located
at the mini Brillouin zone.}}
\end{figure*}

Here, based on the framework discussed in Ref.\cite{EJMele},
we theoretically analyze the emergence of atomic frustrated states
in diatomic molecules, made by two impurity atoms, and hosted by the TBG system.
{Our proposal of atomic frustration is: it represents a molecular-type behavior in which the correlated constituent atoms of the molecule emulate themselves as isolated from each other. Thus, it characterizes a state that locally, i.e., at the atom sites, seems to be noninteracting, while nonlocally still shows, counterintuitively, a correlated behavior by means of an inter-atomic electronic scattering through the molecular host. In equivalent words, the atomic frustration occurs in a molecule that has the trend to split itself into independent atoms, but it is highly prevented of achieving that by its surrounding environment, which should be of Weyl-metallic nature to emerge\cite{Frustration}. It means that just a pair of Dirac cones, with or without time-reversal symmetry broken, is not enough to provide the requirements to form the atomically frustrated state. We call the attention that these cases correspond to Weyl and Dirac semimetals, respectively, which show a pseudogap at the Fermi energy as their primary fingerprint. In case of breaking time-reversal symmetry, the Dirac point of this pair formed by two coincident Weyl nodes with opposite chiralities, then splits into resolved ones in momenta, but still degenerate in energy. Particularly, this operation preserves, as expected, the system inversion symmetry invariance and gapped features. In contrast, if these Weyl nodes minimally move, keeping or not the same momenta, in opposite directions in energy axis, i.e., showing blue and red shifts, which can be accomplished by breaking the inversion symmetry, the band-structure under this change, distinctly, becomes gapless. For atomic frustration, a metallic finite density of states at the Fermi energy (pseudogap closing) assisted by two Weyl nodes with opposite chiralities lifted in energy, at last, constitutes the cornerstone of the proposed state. Thus, due to this frustration in breaking the chemical bound of this state with atomic-like features, it is termed as a molecular state atomically frustrated.} To this end, we consider
the setup sketched in Fig.\ref{fig:Pic1}(a), in particular, settled
to the well-known first twist magic angle $\theta=\theta_{M}=1.08^{{^{\circ}}}$\cite{EJMele}
and with broken inversion symmetry in the Dirac cones, which reside
in the system band-structure {[}Fig.\ref{fig:Pic1}(e){]}. This scenario,
indeed, approaches the corresponding present in the Weyl metals, as
demonstrated in Ref.\cite{Frustration}. In the latter,
a crossover can be observed from the ordinary molecular bonding and
antibonding states, typically from the semimetallic Dirac phase\cite{Dirac,Weyl},
changing to atomic frustrated ones, upon breaking the inversion symmetry. Consequently, this characteristic
then allows the system to enter into the Weyl metal phase\cite{Frustration}.
Simultaneously, atomic frustrated states are revealed as very intriguing, once locally they seem to behave
as those from noninteracting atoms, but counterintuitively, nonlocally
they are still bounded as molecular states. Thus, distinctly from the metallic
Weyl counterparts, atomic frustrated states in our proposal, have the
advantage of being completely tunable, in particular, by twist angles and opposite
gate voltages applied to the graphene monolayers. These gate voltages are indeed, employed
just to break the inversion symmetry in the Dirac cones and close
the pseudogap of the system, in order to produce a metallic phase {[}Fig.\ref{fig:Pic1}(e){]},
where the density of states is finite at the Fermi level. By placing
the twist at the first magic angle condition, in parallel to the pseudogap
closing mechanism previously reported in Weyl metals, here we observe
that the molecular levels suffer strong renormalization characterized
by blue and red shifts, respectively for levels below and above the
Fermi energy. Thus, upon increasing the gate voltages, in the case
of the on-site Coulomb interactions present at the impurity atoms, the graphene Moiré
superlattice then mediate the connection between these dimer's atoms,
just in order to glue them exclusively through itself and build the molecule.
As aftermath, the aforementioned frequency dependent shifts in these
levels, lead to a collapse of the ordinary molecular states into a
single peak {[}Fig.\ref{fig:Pic1}(g){]}. This merging occurs at the
Fermi energy (zero mode) for particle-hole symmetric molecules and
exhibits the atomic frustration character. {We clarify that we designate our impurity dimer by ``particle-hole symmetric molecule'' in the entire text, just to label the regime where the molecular energy levels become symmetrical to the Fermi level $(\omega=0)$. By introducing the molecule with charge mirror symmetry, we report here a phenomenon nonexistent in Weyl metals\cite{Frustration}, which arise from Dirac semimetals and exhibit broken inversion symmetry. Particularly, the TBG nearly flat band Physics, characterized by an extremely low Fermi velocity, then rules the formation of a zero mode atomically frustrated, specially at a sweet spot, namely, the one given by the magic angle. For such a special band-structure, and consequently, different from Weyl metals\cite{Frustration}, where flat bands are lacking, here the system self-energy affected by the twist, strongly renormalizes the energy spectrum of the impurity atoms, leading to particle-hole symmetric zero mode. As aftermath, this atomically frustrated state rises, specially due to its particle-hole symmetry characteristic, as a promising candidate to a Majorana bound state (MBS), which has topological protection\cite{Majorana1,Majorana2}. As MBSs are considered the holy grail for the realization of perfect fault-tolerant topological quantum computing\cite{Majorana1,Majorana2}, the emergence of zero modes in TBG with atomic dimers at magic angle, is expected to stimulate more exploration and future works covering this subject.} Therefore, our results introduce the twistronics of the bilayer graphene system, as an approach to perform unusual molecular bindings in condensed matter platforms.

\section{The Model}

The Hamiltonian of the TBG system, within the theoretical framework
of Ref.\cite{EJMele} and sketched in Figs.\ref{fig:Pic1}(a)-(b),
then follows the Anderson-like Hamiltonian\cite{Anderson1961}, i.e.,
\begin{align}
\mathcal{H} & =\sum_{\text{\textbf{k}}\xi\sigma}\psi_{\text{\textbf{k}}\xi\sigma}^{\dagger}H_{\text{{TBG}}}(\textbf{k}+\textbf{K}_{\xi})\psi_{\text{\textbf{k}}\xi\sigma}+\sum_{j\sigma}\varepsilon_{dj\sigma}d_{j\sigma}^{\dagger}d_{j\sigma}\nonumber \\
 & +\sum_{j}U_{j}d_{j\uparrow}^{\dagger}d_{j\uparrow}d_{j\downarrow}^{\dagger}d_{j\downarrow}+V_{AB}\sum_{j\sigma}(f_{0\sigma}^{\dagger}d_{j\sigma}+\text{H.c.}).\label{eq:Hamiltonian}
\end{align}
The first term in Eq.(\ref{eq:Hamiltonian}) is the low energy Hamiltonian
$H_{\text{{TBG}}}(\textbf{k}+\textbf{K}_{\xi})=\hbar\tilde{v}_{F}(\xi\sigma_{x},\sigma_{y})\cdot\textbf{k}+\xi\sigma_{0}V$
for the pristine TBG, with $\hbar$ as the reduced Planck's constant
and valley indexes $\xi=\pm1$ for the Dirac points {$\overline{\textbf{K}}=\textbf{K}_{\xi=-1}$}
and {$\overline{\textbf{K}'}=\textbf{K}_{\xi=+1}$} at the corners
of the mini Brillouin zone depicted in Figs.\ref{fig:Pic1}(c)-(f).
The first part of $H_{\text{{TBG}}}(\textbf{k}+\textbf{K}_{\xi})$
describes the setup within the length-scale of the Moiré superlattice,
which is the result from the interference pattern due to the misalignment
$\theta$ between the graphene monolayers. {Particularly, the emergent hexagonal
lattice in real space, i.e., the Moiré superlattice, is constituted by the AB and BA spots, where distinct stacking regions made by clusters of carbon atoms appear depicted by green and purple delimiting spheres, respectively in Figs.\ref{fig:Pic1}(a)-(b). As such stacking spots resemble the sites where carbon atoms sit at sublattices A and B of graphene monolayer, thus constituting the basis for the unit cell, these AB and BA regions do similarly, but for the supercell from the Moiré superlattice, which relies on a much larger length-scale. Additionally, as Wannier orbitals are constructed around carbon atoms at sublattices A and B in single graphene to perform the model tight-binding analysis, analogously, corresponding orbitals are also taken into account for the TBG system, but with respect to the AB and BA centers of the Moiré superlattice\cite{EJMele}. Thus, following the employment of the prefix ``super'' in this context, we then make the extrapolation of such a concept to AB and BA spots, where Wannier orbitals are centered, in such a way that hereafter, for a sake of simplicity, we are free to label them as superatoms, in contrast to ordinary carbon atoms from the unit cell of a single graphene sheet, which differently, has a much smaller length-scale.} The superlattice in real space
is characterized by the primitive unit vectors {$\textbf{L}_{1}^{\text{M}}$}
and {$\textbf{L}_{2}^{\text{M}}$}, which define the twist angle-dependent
unit cell area $\Omega_{\theta}=|{\textbf{L}_{1}^{\text{M}}}\times{\textbf{L}_{2}^{\text{M}}}|=\frac{\sqrt{3}a^{2}}{8(\text{sin}\frac{\theta}{2})^{2}}$
\cite{CastroNetoPRB,Review,CastroNetoPRL,EJMele}. The lattice model
then corresponds to an effective graphene monolayer, with renormalized
Fermi velocity given by
\begin{equation}
\tilde{v}_{F}=\frac{v_{F}\left(\hbar^{2}v_{F}^{2}G_{\theta}^{2}-3w_{AB}^{2}\right)}{3w_{AA}^{2}+3w_{AB}^{2}+\hbar^{2}v_{F}^{2}G_{\theta}^{2}},\label{eq:RFV}
\end{equation}
which according to band-structure calculations reveal the existence
of linear dispersion relations at the corners {$\overline{\textbf{K}}$}
and {$\overline{\textbf{K}'}$}, thus giving support to our adopted
theoretical framework\cite{EJMele}. For $\tilde{v}_{F},$ we have
$v_{F}\sim\frac{c}{300}$ the bare Fermi velocity for the graphene
monolayer ($c$ is the speed of light) and $G_{\theta}=\frac{8\pi\text{sin}\frac{\theta}{2}}{3a}$
the distance in momentum space between the aforementioned Dirac points.
Particularly in Fig.\ref{fig:Pic1}(c) {$\overline{\textbf{K}}=\overline{\textbf{K}'}$},
we then have the untwist system with the AA stacking, which corresponds to a metallic-type structure\cite{Review}.
{The AA stacking bilayer arrangement is ideal for the emergence of the atomic frustrated states, once metallic band-structure features from this system, resemble those from Weyl metals\cite{Frustration}. However, the former is energetically unlikely to engineer from a genuine point of view\cite{Review}. Astonishingly, it is still feasible, if we perform a twist and break the inversion symmetry, in particular, in the Bernal-type stacking, which is instead, energetically favorable. This appears in Fig.\ref{fig:Pic1}(e) and is in perfect agreement with Ref.\cite{CastroNetoPRL}, where the authors show that the electronic parabolic dispersion of the Bernal stacking\cite{Review}, surprisingly, then turns into linear under twist and without inversion symmetry, just by applying opposite gate voltages to graphene sheets. It means that our findings do not depend on the AA stacking to manifest. Hence, our choice to show its band-structure at Fig.\ref{fig:Pic1}(c), just highlights that this ideal metallic phase is entirely accomplishable by bilayer graphene twistronics and gate microscopy, since we begin with the bilayer graphene in Bernal stacking\cite{CastroNetoPRL}.}
The lattice parameter for the graphene monolayer is $a=\frac{\hbar v_{F}}{2135.4\text{{meV}}},$
$w_{AB}=97.5\text{{meV}}$ and $w_{AA}=79.7\text{{meV}}$ are the
AB and AA hopping terms, respectively and $\sigma_{x},\sigma_{y}$
the Pauli matrices. The first twist magic angle condition is obtained
from $\hbar^{2}v_{F}^{2}G_{\theta}^{2}=3w_{AB}^{2}$ in Eq.(\ref{eq:RFV}),
once it leads to the flat band characterized by $\tilde{v}_{F}=0,$
giving rise to $\theta=\theta_{M}=1.08^{{^{\circ}}}.$ Then, $H_{\text{{TBG}}}(\textbf{k}+\textbf{K}_{\xi})$
is the momentum representation of the tight-binding model with Wannier
orbitals centered at the AB and BA superatoms\cite{EJMele}, characterized by
the spinor $\psi_{\text{\textbf{k}}\xi\sigma}^{\dagger}=(\begin{array}{cc}
\psi_{AB\text{\textbf{k}}\xi\sigma}^{\dagger} & \psi_{BA\text{\textbf{k}}\xi\sigma}^{\dagger}\end{array})$ describing the electronic states by the operators $\psi_{AB\text{\textbf{k}}\xi\sigma}$
and $\psi_{BA\text{\textbf{k}}\xi\sigma},$ with wave vector $\bold k$,
valley index $\xi$ and spin $\sigma.$ Additionally, the term $\xi\sigma_{0}V$
in $H_{\text{{TBG}}}(\textbf{k}+\textbf{K}_{\xi}),$ with $\sigma_{0}$
being the identity matrix, is put to break the inversion symmetry
of the Dirac cones. This breaking can be performed by applying opposite
top and bottom gate voltages $\pm V$ to the graphene monolayers.
Such a demonstration can be found in Refs.\cite{CastroNetoPRL,EJMele}
and leads to the closing of the system pseudogap, driving the TBG
arrangement to a metallic phase {[}Fig.\ref{fig:Pic1}(e){]}. We shall
see that atomic frustrated states become feasible in such a regime,
as demonstrated for Weyl metals\cite{Frustration}, where
the chirality degree of freedom plays the role of the current valley
index. The operators $d_{j\sigma}^{\dagger},d_{j\sigma}$ describe
the electronic states of the individual impurity atoms ($j=1,2$)
(red spheres within the AB site depicted in Figs.\ref{fig:Pic1}(a)-(b))
with single-particle energies $\varepsilon_{dj\sigma}$ and the on-site
Coulomb correlation energies $U_{j}$. The term $V_{AB}=v_{0}$ is the
coupling strength between the AB site and the dimer's atoms. We would
like to highlight that we extrapolate the use of the concept of the
dimer word, which here still represents the diatomic molecule itself,
even in the absence of a direct hopping between its constituent atoms,
as it occurs in the current theoretical framework. Contrasting with
the standard molecular bindings, which are defined in terms of this
tunneling term, we then reveal that the considered impurity atoms
of the system, distinctly, connect to each other through the host,
in particular, via the AB site of the superlattice depicted in Figs.\ref{fig:Pic1}(a)-(b).
{It is worth mentioning that we strictly follow both Ref.\cite{EJMele} and
its Supplementary Material, in which the grounds for sustaining impurities placed at AB sites (or BA equivalently) in Moiré superlattice, appear well-established. Indeed, such impurities differ completely from those ordinary placed above isolated carbon atoms in single graphene sheet. According to tight-binding calculations, in particular for low twist angles and energies, the system Wannier orbitals around AB and BA sites, do not exhibit maximum localization on themselves. Instead, a three-peak localization at the triangle corners made by neighboring hexagonal centers (hollow sites) is revealed, as depicted in Figs.2(b)-(c) from Ref.\cite{EJMele}. Hence, we should be aware that impurities with sizes of carbon atoms atop these hollow centers, will overlap on equal footing, inevitably, to both Wannier orbitals. However, in order to catch preferably just one set of these wave-functions, in such a way that AB sites become energetically favorable to host impurities, one should, as pointed out in Supplementary Material from Ref.\cite{EJMele}, engineer impurities bigger than carbon atoms, but still within the Moiré superlattice length-scale. Thus, according to Figs.2(b)-(c) from Ref.\cite{EJMele}, by sitting large impurities at one AB site, their wave-functions overlap stronger with this spot, where the Wannier orbital has a higher amplitude, with respect to Wannier orbitals from the nearest three surrounding BA sites, which at this AB site, show instead, a lower amplitude and then, overlap with the impurities, as aftermath, in a much weaker strength. By this manner, we can safely describe the hybridization between the AB site and impurities by the fermionic field operator given by
\begin{equation}
f_{0\sigma}=\sum_{\text{\textbf{k}}\xi}\psi_{AB\text{\textbf{k}}\xi\sigma}.\label{eq:f0}
\end{equation}
Under these assumptions, we propose that these two large impurities of the dimer, could be designed by making atomic clusters, similarly to Ref.\cite{Cluster}. By this reason, we expect that this experimental approach can engineer, in the future, proper impurities for the Moiré superlattice in TBG, provided that a given chemical element of such, is correctly considered. Hence, such an approach in Ref.\cite{Cluster} rises as promising and deserves extra exploration for the realization of our current theoretical proposal. Particularly, the particle-hole symmetric regime, will be crucial to define zero modes, as we will see later on, as atomic frustrated states. Thus, the experimental challenge consists of extrapolating the technique to TBG system, taking into account cluster sizes now bigger than carbon atoms and within the Moiré superlattice length-scale.}
This assumption can be safely adopted, once in the length-scale of the graphene monolayers the impurity
atoms are assumed to be far apart, in such a way that the overlapping
between the wave functions from this pair of atoms vanishes. However,
in the Moiré superlattice length-scale, these impurities are close
enough and locally coupled to the AB site via $f_{0\sigma}$. It means
that as the impurity atoms perceive the superlattice as their host
instead of the graphene monolayers individually, we are still free
to call this pair of impurities by dimer's atoms, once this dimer
stays well-established by the superlattice environment. Thus, we can
solely focus on this inter-atomic mediation afforded by the Moiré
superlattice, which is indeed the cornerstone to provide molecular
bindings in the TBG system. Additionally, we shall see later on that
such a host-assisted molecular binding, as a matter of fact, will be of capital relevance
to analyze the molecular correlation function.

The spectral features of the dimer's atoms can be extracted from their
Density of States (DOS) $\rho_{jj}^{\sigma}(\omega)=-\frac{1}{\pi}\text{{Im}}[\tilde{\mathcal{G}}_{j\sigma|j\sigma}(\omega)]$
($j=1,2$), where $\tilde{\mathcal{G}}_{j\sigma|j\sigma}(\omega)$
is the Green's function (GF) in energy domain $\omega.$ This
DOS, via differential conductance measurements, can be detected by
an STM-tip (\textit{Scanning Tunneling Microscope}-tip)\cite{STM1,STM2}
placed above the $\text{{j}}^{\text{{th}}}$ atom of the dimer. To
obtain such, we first define this quantity in time domain as $\mathcal{G}_{j\sigma|j\sigma}(t)=-\frac{i}{\hbar}\theta(t)<\{d_{j\sigma}(t),d_{j\sigma}^{\dagger}(0)\}>_{\mathcal{H}}$
and later on, apply to it the time-Fourier transform, in particular
with the Hubbard-I approximation\cite{Hubbard1963} [see the Appendices],
which gives
\begin{eqnarray}
\tilde{\mathcal{G}}_{j\sigma|j\sigma}(\omega) & = & \frac{\lambda_{j}^{\bar{\sigma}}}{g_{j\sigma|j\sigma}^{-1}(\omega)-\lambda_{j}^{\bar{\sigma}}{{\Sigma}_{0}(\omega)g_{l\sigma|l\sigma}(\omega)\lambda_{l}^{\bar{\sigma}}{\Sigma}_{0}(\omega)}},\nonumber \\
\label{eq:Gjj}
\end{eqnarray}
where $\bar{\sigma}=-\sigma$, $j\neq l$,
\begin{equation}
g_{j\sigma|j\sigma}(\omega)=\frac{1}{\omega-\varepsilon_{dj\sigma}-{\Sigma}_{0}(\omega)}
\end{equation}
is the single-atom noninteracting GF,
\begin{equation}
\lambda_{j}^{\sigma}=1+\frac{U_{j}}{g_{j\bar{\sigma}|j\bar{\sigma}}^{-1}(\omega)-U_{j}}\bigl\langle n_{j\sigma}\bigr\rangle\label{eq:Lambda}
\end{equation}
is the spectral weight,
\begin{equation}
\bigl\langle n_{j\sigma}\bigr\rangle=-\frac{1}{\pi}\int_{-D}^{+D}n_{_{F}}(\omega)\textrm{Im}[\mathcal{\tilde{G}}_{j\sigma|j\sigma}(\omega)]d\omega\label{eq:Number}
\end{equation}
is the atom occupation, $D$ is the energy cutoff and ${\Sigma}_{0}(\omega)=\sum_{\xi}\Sigma_{0}^{\xi}(\omega)$
the self-energy given by
\begin{align}
\Sigma_{0}^{\xi}(\omega)=\frac{v_{0}^{2}\Omega_{\theta}}{4\pi\hbar^{2}\tilde{v}_{F}^{2}}(\omega_{\xi}\text{ln}\frac{\omega_{\xi}^{2}}{|D^{2}-\omega_{\xi}^{2}|}-i\pi|\omega_{\xi}|) & ,\label{eq:SE0}
\end{align}
with $\omega_{\xi}=\omega-\xi V.$ We call attention that $-\text{{Im}}{\Sigma}_{0}(\omega)$
contains the Moiré superlattice density of states with a pseudogap
at the Fermi level $\omega=0$ and $V=0.$ However, due to $V\neq0$
in $\omega_{\xi}$, the $-\text{{Im}}\Sigma_{0}^{\xi}(\omega)$ and
$\text{{Re}}\Sigma_{0}^{\xi}(\omega)$ quantities, as functions of $\omega$
for the valley indexes $\xi=+1$ and $\xi=-1,$ show blue and red
shifts in their profiles, respectively. Particularly with $V\neq0,$
the pseudogap closes in $-\text{{Im}}{\Sigma}_{0}(\omega)$ and a
metallic plateau takes place, with a finite density of states around
the Fermi energy. Concerning the explicit form of Eq.(\ref{eq:Gjj}),
we can predict in our system, the appearance of bonding and antibonding
molecular states. Before to perceive how they emerge, let us assume
the impurity atoms decoupled from each other, which corresponds to impose $\lambda_{j}^{\bar{\sigma}}{{\Sigma}_{0}(\omega)g_{l\sigma|l\sigma}(\omega)\lambda_{l}^{\bar{\sigma}}{\Sigma}_{0}(\omega)}=0$
in Eq.(\ref{eq:Gjj}), just in order to reveal spectrally the decoupled
behavior {[}$\tilde{\mathcal{G}}_{j\sigma|j\sigma}^{\text{{dec}}}(\omega)${]} of such atoms in $\rho_{jj}^{\sigma}(\omega)$ ($j=1,2$).
Under this assumption, we recover the well-established GF for the
Hubbard-I approximation in the single atom problem\cite{Hubbard1963}:
\begin{equation}
\tilde{\mathcal{G}}_{j\sigma|j\sigma}^{\text{{dec}}}(\omega)=\frac{1-\bigl\langle n_{j\bar{\sigma}}\bigr\rangle}{\omega-\varepsilon_{dj\sigma}-{\Sigma}_{0}(\omega)}+\frac{\bigl\langle n_{j\bar{\sigma}}\bigr\rangle}{\omega-\varepsilon_{dj\sigma}-U_{j}-{\Sigma}_{0}(\omega)},\label{eq:HB_I}
\end{equation}
where we recognize the two bare (atomic) poles $\varepsilon_{dj\sigma}<0$
and $\varepsilon_{dj\sigma}+U_{j}>0,$ which in the presence of an
electronic reservoir, namely, our Moiré superlattice, are converted
into bands, with peak positions and widths being renormalized and broadened by $\text{{Re}}{\Sigma}_{0}(\omega)$
and $-\text{{Im}}{\Sigma}_{0}(\omega),$ respectively. {Particularly for large twist angles $\theta>>\theta_M$, let us say for instance $\theta=13^{{^{\circ}}}$ and $\theta=30^{{^{\circ}}}$, just to illustrate our goal here, we astonishingly reveal $\tilde{v}_{F}\rightarrow{v}_{F}$ and ${\Sigma}_{0}(\omega)\rightarrow0$ from Eqs.(\ref{eq:RFV}) and (\ref{eq:SE0}), respectively. This turns out when the interlayer coupling for graphene monolayers approaches ideal suppression, once the fingerprint of such relies on recovering the pristine Fermi velocity ${v}_{F}$\cite{Review}. Consequently, the Moiré superlattice-impurity hybrid system picture of Fig.\ref{fig:Pic1}(a) breaks down and impurities show Dirac-delta like-behavior, i.e., $\rho_{jj}(\omega)\approx\delta(\omega-\varepsilon_{dj\sigma})+\delta(\omega-\varepsilon_{dj\sigma}-U)$, typically for isolated atomic states. In this regime, twist angles and gate voltages, as expected, play no role to sustain molecular bindings, clearly due to the scenario very close to perfect decoupled impurities and isolated graphene sheets.} By turning-on the inter-atomic correlation by keeping the term $\lambda_{j}^{\bar{\sigma}}{{\Sigma}_{0}(\omega)g_{l\sigma|l\sigma}(\omega)\lambda_{l}^{\bar{\sigma}}{\Sigma}_{0}(\omega)}\neq0$
in Eq.(\ref{eq:Gjj}), the resonant level $\varepsilon_{dj\sigma}<0$
($\varepsilon_{dj\sigma}+U_{j}>0$) splits into molecular levels,
known as bonding and antibonding states. These states are still ruled
by the self-energy ${\Sigma}_{0}(\omega),$ as we shall see.
It is worth mentioning that our central finding relies on such a quantity,
namely, the self-energy ${\Sigma}_{0}(\omega),$ in particular at
$\theta=\theta_{M}=1.08^{{^{\circ}}}$ and with $V\neq0,$ together
with the parameter $\lambda_{j}^{\bar{\sigma}},$ which is dependent
on the on-site Coulomb interaction $U_{j}.$ In this case, when the
particle-hole $2\varepsilon_{dj\sigma}+U_{j}=0$ constraint is fulfilled,
the splittings of the Hubbard bands around $\varepsilon_{dj\sigma}<0$
and $\varepsilon_{dj\sigma}+U_{j}>0,$ then become suppressed in $\rho_{jj}^{\sigma}(\omega)$
($j=1,2$) upon increasing $V,$ as illustrated in Fig.\ref{fig:Pic1}(g),
with a single peak at the Fermi level as a zero mode. This occurs, as we shall see, due to the strong renormalization
of the molecular levels arising from $\text{{Re}}{\Sigma}_{0}(\omega),$
which leads to blue and red shifts in the split bands
$\varepsilon_{dj\sigma}<0$ and $\varepsilon_{dj\sigma}+U_{j}>0,$
respectively. Thereafter, the molecular levels collapse into a single
one {[}Fig.\ref{fig:Pic1}(g){]}. {This can be achieved by choosing $2\varepsilon_{dj\sigma}+U_{j}=0$ in the model as stated previously, and verified in its spectral analysis by $\rho_{jj}^{\sigma}(\omega)=\rho_{jj}^{\sigma}(-\omega)$. Such even-parity feature is a straight consequence of this particular Hamiltonian, which is invariant under a standard particle-hole transformation\cite{Kondo}, characterized by the half-filling occupation $\bigl\langle n_{j\bar{\sigma}}\bigr\rangle=1/2$. Thus, due to this charge mirror symmetry, we naturally adopt the term ``particle-hole symmetric molecule'' for the dimer's atoms. However, if the BG system from the experimental framework, instead of being prepared in suspended geometry and within a high-vacuum chamber, upon considering, the typical $\text{SiC}$ substrate, as the BG platform, it can introduce a doping level in the entire system spectrum\cite{Puddles}. As aftermath, it will compromise the particle-hole symmetry, due to the lifting of the Dirac point and Fermi level degeneracy\cite{Puddles}, thus yielding $\bigl\langle n_{j\bar{\sigma}}\bigr\rangle\neq1/2$ and a single peak state off the zero mode. Despite such a contamination, the recovery of this degeneracy becomes feasible, just by controlling nitrogen dopants on $\text{SiC}$. As a result, these dopants lead to electron-lack puddles in graphene, thus decreasing the original doping by approaching the Fermi level towards the Dirac point\cite{Puddles}. As pointed out in Ref.\cite{Doping}, instead of doping the substrate as in Ref.\cite{Puddles}, n-type or p-type doping techniques performed directly in graphene, similarly, could rule the system particle-hole symmetry. Thus, for a given doping and slight twist, the $2\varepsilon_{dj\sigma}+U_{j}=0$ situation could be, in principle, achieved as well, giving rise to a zero mode at magic angle.} Intuitively, with this single peak,
we naturally infer that the molecule is dissociated, once the splittings
of the atomic levels are absent in $\rho_{jj}^{\sigma}(\omega)$ ($j=1,2$).
However, in order to put such an usual statement to the test, we should
focus on the molecular correlation function $\delta\rho_{jl}(\omega),$
namely, $\sum_{jl}\delta\rho_{jl}(\omega)=-\frac{1}{\pi}\text{{Im}}[\tilde{\mathcal{G}}^{\text{{corr}}}(\omega)]$
($j\neq l=1,2$), which is determined from the time-Fourier transform
of the GF $\mathcal{G}^{\text{{corr}}}(t)=-\frac{i}{\hbar}\theta(t)<\{f_{0\sigma}(t),f_{0\sigma}^{\dagger}(0)\}>_{\mathcal{H}}.$
As a result, we find
\begin{align}
\delta\rho_{jl}(\omega) & =-\frac{1}{\pi v_{0}^{2}}\sum_{\xi\xi'\sigma}\textrm{Im}[\Sigma_{0}^{\xi}(\omega){\cal {\cal \tilde{G}}}_{j\sigma|l\sigma}(\omega)\Sigma_{0}^{\xi'}(\omega)],\label{eq:LDOSjl}
\end{align}
expressed in terms of the GF
\begin{eqnarray}
\tilde{{\cal G}}_{{j\sigma}|{{l}\sigma}}(\omega) & = & g_{j\sigma|j\sigma}(\omega){\lambda_{j}^{\bar{\sigma}}{\Sigma}_{0}(\omega)}{\cal \tilde{{\cal G}}}_{{{l}\sigma}|{{l}\sigma}}(\omega),\label{eq:Gjl}
\end{eqnarray}
which in time domain is defined by $\mathcal{G}_{j\sigma|l\sigma}(t)=-\frac{i}{\hbar}\theta(t)<\{d_{j\sigma}(t),d_{l\sigma}^{\dagger}(0)\}>_{\mathcal{H}}$
and accounts for the interference processes between the atoms $j$
and $l$ $(j\neq l)$ through the Moiré superlattice, in particular,
assisted by the on-site Coulomb correlation $U_{j}$ within the parameter
$\lambda_{j}^{\bar{\sigma}}.$ Thus, if nonlocally $\delta\rho_{jl}(\omega)$
is finite for $j\neq l$, despite the local absence of splittings
in the atomic levels for $\rho_{jj}^{\sigma}(\omega)$ ($j=1,2$),
we will see that the dimer's atoms, then look like as free atoms individually,
but they still build the molecule. In this scenario, we find an unusual
molecular binding, which we introduce as atomically frustrated.

\section{Results and Discussion}

In what follows, we adopt model parameters obeying the particle-hole
symmetric condition $2\varepsilon_{dj\sigma}+U_{j}=0,$ by considering
$\varepsilon_{dj\sigma}=-0.01D$ with $D=1\text{{eV}}$ (the energy
cutoff) and $v_{0}=0.1\text{{meV}}.$ Naturally, once the region above
the Fermi level is the mirror for the corresponding below, the profiles
presentation for $\rho_{jj}^{\sigma}(\omega)$ ($j=1,2$) in this
domain is needless. It means that the Hubbard bands around $\varepsilon_{dj\sigma}<0$
and $\varepsilon_{dj\sigma}+U_{j}>0$ are equally distant from the
Fermi energy and exhibit the same spectral profiles.

\begin{figure*}[ht]
\centering\includegraphics[width=0.9\textwidth,height=0.5\textheight]{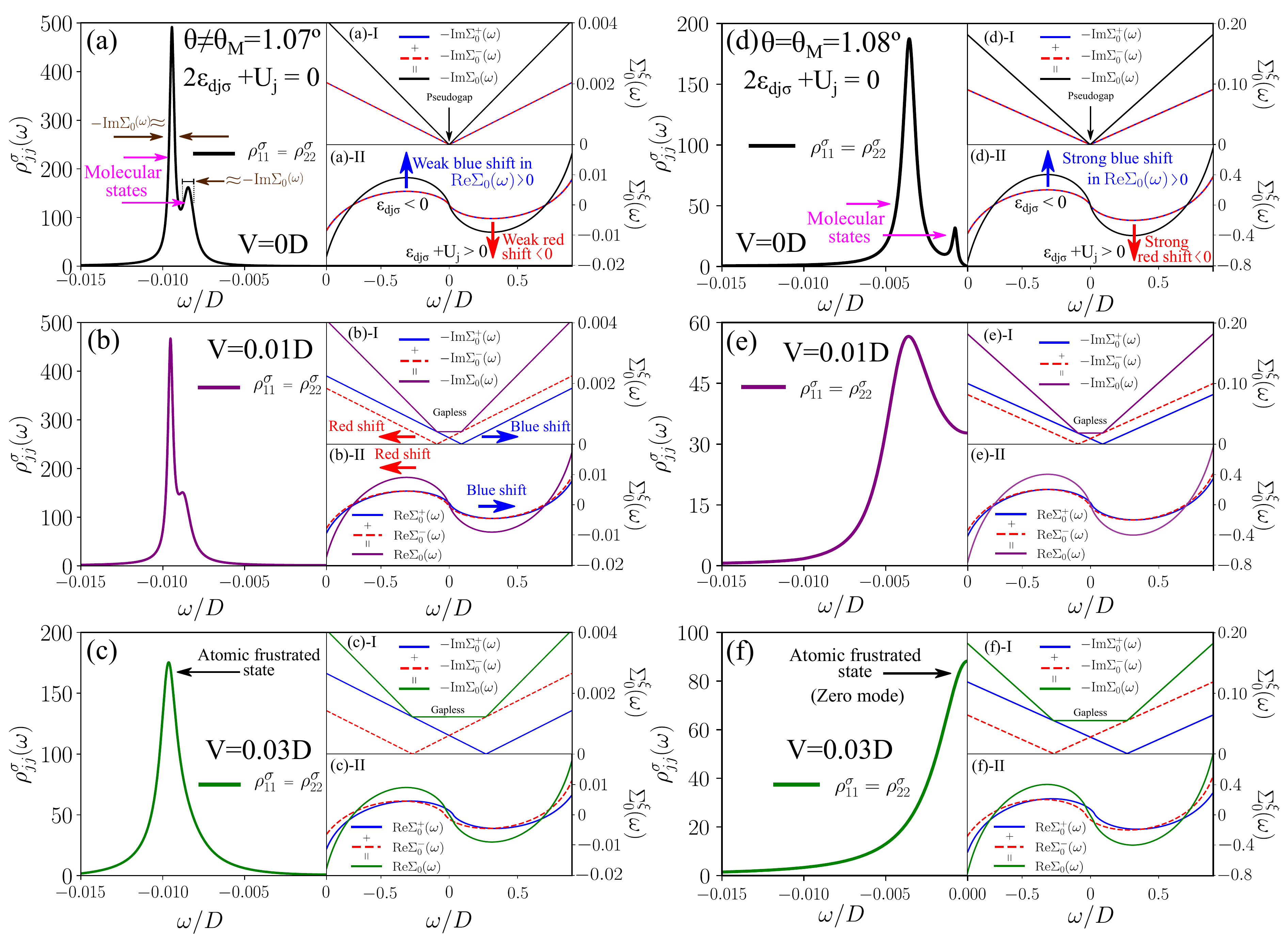}\caption{\label{fig:Pic2} {(Color online) Panels (a)-(c): Crossover observed
in the spectral density of states $\rho_{jj}^{\sigma}(\omega)$ ($j=1,2$) for the dimer's atoms, with $\varepsilon_{dj\sigma}=-0.01D$ and particle-hole symmetry ($2\varepsilon_{dj\sigma}+U_{j}=0$), as a function of the single particle energy $\omega$ in the negative domain, in particular, off the first twist magic angle, i.e., $\theta=1.07^{{^{\circ}}}$. Here, we visualize the evolution from the ordinary
molecular states (level splitting in $\varepsilon_{dj\sigma}$
into bonding and antibonding states) at panel (a), which are approximately
broadened by $-\text{{Im}}\Sigma_{0}(\omega)$, via the breaking down
of the inversion symmetry by the gate voltage $V$ {[}panels (b)
and (c){]}, towards the final atomic frustrated state, where the aforementioned level splitting
is absent, as depicted at panel (c). Additionally, a second
atomic frustrated state exists in the positive energy domain, but it is not shown,
due to the aforementioned charge mirror symmetry for the dimer. Notice
that the peak in panel (c) is the result of the merge between
the molecular peaks from panels (a)-(b), ruled by the pseudogap closing
in $-\text{{Im}}\Sigma_{0}(\omega),$ which is characterized by a metallic
(gapless) plateau {[}inset panels (b)-I and (c)-I{]}. The plateau
amplitude defines the broadening of the molecular peaks and arises
from the blue and red shifts in the profiles for $-\text{{Im}}\Sigma_{0}^{+}(\omega)$
and $-\text{{Im}}\Sigma_{0}^{-}(\omega),$ respectively. Consequently,
the molecular states collapse into a unique resonant state, as in
panel (c), but still placed around $\varepsilon_{dj\sigma}=-0.01D$,
once the blue and red shifts in $\text{{Re}}\Sigma_{0}(\omega)$ are
in the weak regime (up to $\approx10^{-2}D$ according to the right
axis of the inset panels (a)-II, (b)-II and (c)-II). Such a single
peak is still considered of molecular nature, but with atomic frustration
as in the metallic Weyl counterparts\cite{Frustration}. Panels (d)-(f):
The corresponding crossover for panels (a)-(c) in the case of the
first twist magic angle $\theta=\theta_{M}=1.08^{{^{\circ}}}.$ Distinctly
from $\theta=1.07^{{^{\circ}}},$ beyond the pseudogap closing mechanism
in $-\text{{Im}}\Sigma_{0}(\omega)$ {[}panels (e)-I and (f)-I{]},
the blue and red shifts in $\text{{Re}}\Sigma_{0}(\omega)$ {[}panels
(d)-II, (e)-II and (f)-II{]} enter into the strong regime (up to $\approx10^{-1}D$,
i.e., 10 times the $\theta=1.07^{{^{\circ}}}$ case), thus also ruling
the emergence of the atomic frustrated state. This second mechanism
does not manifest in Weyl metals, once the Fermi velocity is not tunable
by twist angles. By this manner, the molecular states at negative
and positive (not shown) energy positions merge exactly at $\omega=0,$
therefore constituting a molecular zero mode of energy {[}panel (f){]},
which is considered atomically frustrated {[}see Figs.\ref{fig:Pic3}(a)-(c){]},
cf. details in the main text.}}
\end{figure*}

In Fig.\ref{fig:Pic2}, we analyze $\rho_{jj}^{\sigma}(\omega)$ ($j=1,2$)
in two scenarios. The first treats the twist angle $\theta=1.07^{{^{\circ}}}$
{[}Figs.\ref{fig:Pic2}(a)-(c){]}, which is off the magic angle condition,
while the second is exactly settled to such a point, i.e., $\theta=\theta_{M}=1.08^{{^{\circ}}}$
{[}Figs.\ref{fig:Pic2}(d)-(f){]}. Both share an ordinary characteristic
when the inversion symmetry is preserved ($V=0$) in the Dirac cones
for the mini Brillouin zone: the emergence of the molecular bonding
and antibonding states due to the atomic level splitting in $\varepsilon_{dj\sigma}=-0.01D,$
as pointed out by the rose arrows in Figs.\ref{fig:Pic2}(a) and (d).
Although they have this feature in common, these states are broadened
by $-\text{{Im}}\Sigma_{0}(\omega),$ whose resonant peak positions
depend on $\text{{Re}}\Sigma_{0}(\omega),$ both distinctly for each $\theta.$ Particularly, $-\text{{Im}}\Sigma_{0}(\omega)$
shows a semimetallic pseudogap {[}Figs.\ref{fig:Pic2}(a)-I and (d)-I{]}
leading to narrow widths in these states when $\varepsilon_{dj\sigma}$
approaches the Fermi energy, which is the point where this gap opens
with the decreasing in the number of density of states. This narrowing
then allows to resolve the splitting of $\varepsilon_{dj\sigma}=-0.01D$
into the molecular levels present in $\rho_{jj}^{\sigma}(\omega).$
However, the $\text{{Re}}\Sigma_{0}(\omega)$ quantity magnifies differently
for $\theta=1.07^{{^{\circ}}}$ and $\theta=\theta_{M}=1.08^{{^{\circ}}}$
{[}Figs.\ref{fig:Pic2}(a)-II and (d)-II{]} according to the blue
($\text{{Re}}\Sigma_{0}(\omega)>0$) and red ($\text{{Re}}\Sigma_{0}(\omega)<0$)
shifts indicated by the vertical arrows for state levels placed below
($\varepsilon_{dj\sigma}<0$) and above ($\varepsilon_{dj\sigma}+U_{j}>0$)
the Fermi energy ($\omega=0$), respectively. Notice that the renormalization
from $\text{{Re}}\Sigma_{0}(\omega)$ in the levels for $\rho_{jj}^{\sigma}(\omega)$
is stronger when $\theta=\theta_{M}=1.08^{{^{\circ}}},$ once they
are driven closer to the Fermi energy {[}Fig.\ref{fig:Pic2}(d){]}
than in the $\theta=1.07^{{^{\circ}}}$ case, where we still observe
the molecular states around $\varepsilon_{dj\sigma}=-0.01D$ {[}Fig.\ref{fig:Pic2}(a){]}.
\begin{figure*}[ht]
\centering\includegraphics[width=0.82\textwidth,height=0.28\textheight]{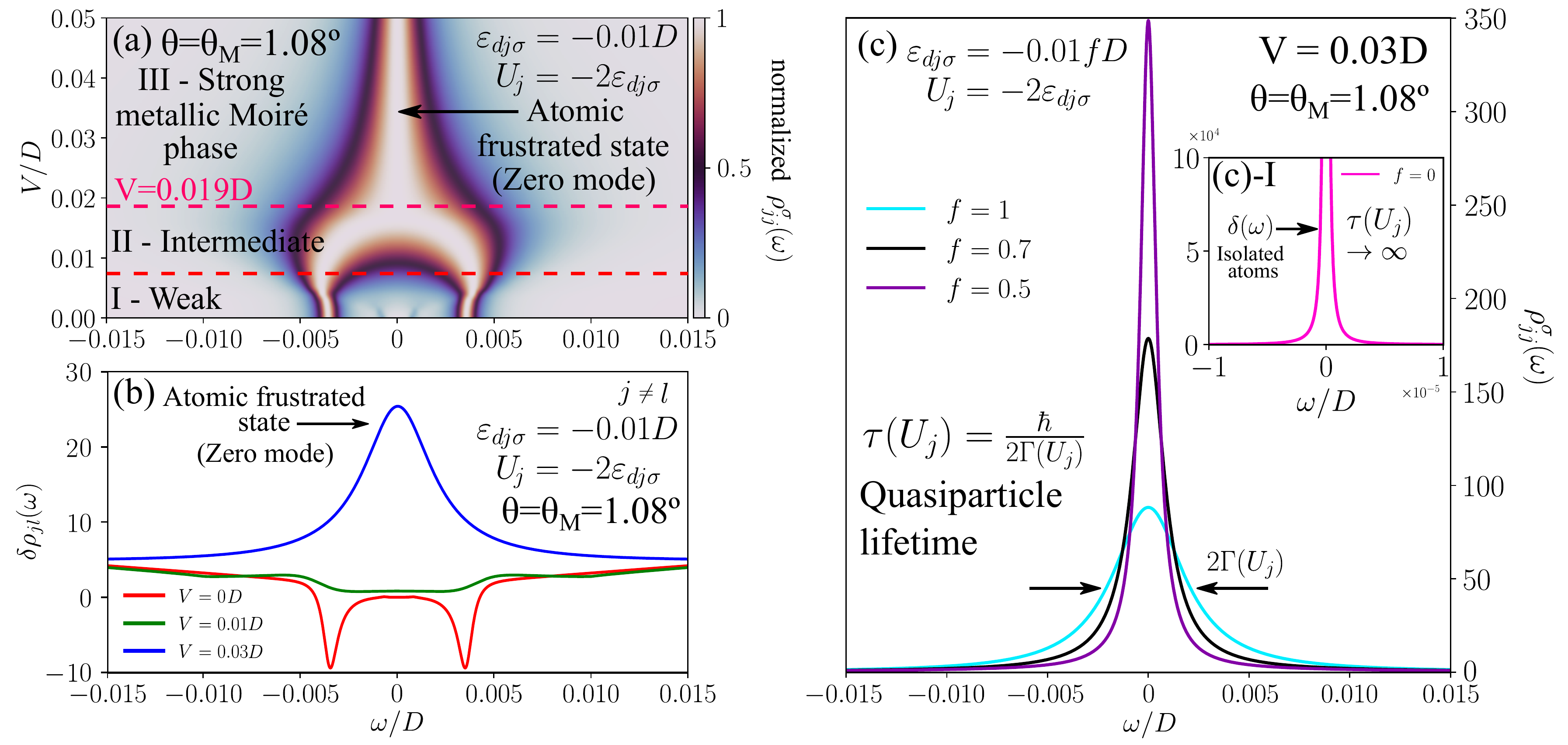}
\caption{\label{fig:Pic3} {(Color online) Panel (a): Color map of the normalized
$\rho_{jj}^{\sigma}(\omega)$ ($j=1,2$) for the dimer's atoms, with $\varepsilon_{dj\sigma}=-0.01D$
and particle-hole symmetry $(U_{j}=-2\varepsilon_{dj\sigma}),$ spanned by $V$ and $\omega$ at the first twist
magic angle $\theta=\theta_{M}=1.08^{{^{\circ}}}.$ A crossover is
shown from the ordinary molecular states (I- the weak metallic Moiré
phase) triggered by the continuous changing of the $V$ parameter
(II- the intermediate metallic Moiré phase), which results in the
levels collapse into a single zero mode (III- the strong metallic
Moiré phase, {with $V=0.019D$ as the zero mode threshold to emerge}). This mode is atomically frustrated, as we shall verify
later on. Panel (b): Molecular correlation function $\delta\rho_{jl}(\omega)$
($j\protect\neq l=1,2$) as a function of $\omega,$ with particle-hole
symmetry $(U_{j}=-2\varepsilon_{dj\sigma}).$ Finite values of $\delta\rho_{jl}(\omega)$
identify molecular bindings as well-established, due to usual splitting
of the atomic level into molecular states, such as the one observed
in Fig.\ref{fig:Pic2}(d) with $V=0.$ However, when this splitting
is absent as in Fig.\ref{fig:Pic2}(f) with $V=0.03D,$ it suggests
the decoupling of the atoms from each other and the recovering of
an atomic state. Counterintuitively, the correlation $\delta\rho_{jl}(\omega)$
is still finite even in such a situation. It means that locally the
dimer's atoms behave, by means of $\rho_{jj}^{\sigma}(\omega),$ as
uncoupled atoms due to the spectral single peak characteristic, while
its nonlocal behavior ($\delta\rho_{jl}(\omega)\protect\neq0,$ $j\protect\neq l$)
is of molecular nature mediated by the Moiré superlattice, in particular,
assisted by the atomic Coulomb interaction (III- strong metallic Moiré
phase). For this case, we call this peculiar molecular mode at $\omega=0$
that seems to be atomic, as a molecular zero mode atomically frustrated.
Panel (c): Coulomb interaction ($U_{j}$) dependence in the particle-hole
symmetric ($\varepsilon_{dj\sigma}=-0.01fD$ and $U_{j}=-2\varepsilon_{dj\sigma}$)
spectral density of states $\rho_{jj}^{\sigma}(\omega)$ ($j=1,2$)
for the atomic frustrated state regime $(V=0.03D).$ We see that the
\textit{quasiparticle} lifetime $\tau(U_{j})=\frac{\hbar}{2\Gamma(U_{j})}$,
which is inversely proportional to the spectral width $2\Gamma(U_{j})$
of $\rho_{jj}^{\sigma}(\omega)$, increases upon decreasing $f.$
In the inset (c)-I, the limit case with $f=0$ $(U_{j}=0)$ leads
to a $\rho_{jj}^{\sigma}(\omega)\approx\delta(\omega)$ behavior.
The latter denotes that the dimer's atoms become completely isolated
from the host, in such a way that the atomic frustrated state, which
should be assisted by the $U_{j}$ term to exist, quenches.}}
\end{figure*}
As $\text{{Re}}\Sigma_{0}(\omega)$ depends on the Fermi velocity
inversely, in particular for $\theta=\theta_{M}=1.08^{{^{\circ}}},$
it increases substantially and displaces strongly with blue (red)
shift the molecular negative (positive) peak positions. It is worth
citing that $-\text{{Im}}\Sigma_{0}^{\xi}(\omega)$ and $\text{{Re}}\Sigma_{0}^{\xi}(\omega)$
for the valley indexes $\xi=\pm1$ of the mini Brillouin zone, as we can see in
the inset panels from Figs.\ref{fig:Pic2}(a) and (d), are degenerate
when the inversion symmetry is preserved. Interestingly enough, by
breaking down such a symmetry with $V\neq0,$ the ordinary molecular
profiles in $\rho_{jj}^{\sigma}(\omega)$ change drastically, once
the degeneracy in $-\text{{Im}}\Sigma_{0}^{\xi}(\omega)$ and $\text{{Re}}\Sigma_{0}^{\xi}(\omega)$
quantities is lifted. Their profiles split into blue and red shifts
for $\xi=+1$ and $\xi=-1,$ respectively {[}inset panels for Figs.\ref{fig:Pic2}(b),
(c), (e) and (f){]}. Particularly in Fig.\ref{fig:Pic2}(b) {[}Fig.\ref{fig:Pic2}(e){]},
we observe partially (completely) collapse of the molecular states
with $V=0.01D.$ The case for Fig.\ref{fig:Pic2}(b) is ruled exclusively
by the blue and red shifts in $-\text{{Im}}\Sigma_{0}^{\xi}(\omega)$
profiles, which close the pseudogap in $-\text{{Im}}\Sigma_{0}(\omega)$
and make the molecular states broader, but not enough for the entire
collapse. This constitutes a gapless phase of the Moiré superlattice,
with a metallic plateau emerging in $-\text{{Im}}\Sigma_{0}(\omega),$
whose amplitude defines the broadening of the molecular peaks. This phenomenon
has been already observed in the metallic Weyl counterparts, triggered
exactly by the pseudogap closing from the semimetallic Dirac phase.
It leads to molecular states atomically frustrated\cite{Frustration},
in particular off the Fermi energy with the increase of $V,$ as depicted by the resonant peak
in Fig.\ref{fig:Pic2}(c) with $V=0.03D.$ We should remember that, there is also another one in the positive energy domain, due to the existing charge mirror symmetry in the dimer's atoms. However, for $\theta=\theta_{M}=1.08^{{^{\circ}}}$
and $V=0.03D,$ the strong regime of the vertical blue and red shifts
in $\text{{Re}}\Sigma_{0}(\omega)$ is enough to allow the complete
merge between the molecular peaks into a single resonant structure,
in particular, exactly at the Fermi energy {[}Fig.\ref{fig:Pic2}(f){]}.
As a matter of fact, the blue and red shifts in $\text{{Re}}\Sigma_{0}(\omega)$
at the inset panels (d)-II, (e)-II and (f)-II from Fig.\ref{fig:Pic2}
are considered pronounced, due to their enhanced amplitudes,
which are up to $\approx10^{-1}D$, i.e., 10 times greater when compared
to the $\theta=1.07^{{^{\circ}}}$ case shown in the inset panels
(a)-II, (b)-II and (c)-II of the same figure. Thus, distinctly from
the Weyl metal case\cite{Frustration}, this single peak still has its origin
from the pseudogap closing, but in addition to such a mechanism, in
order to become a zero mode, it needs also the one from the strong
renormalization of the molecular levels, due to the $\text{{Re}}\Sigma_{0}(\omega)$
function. Such a feature then allows the blue and red shifts for the
molecular states found below and above the Fermi energy, respectively
and inevitably, to a collapsed final state wherein the ordinary molecular
peaks are not resolved as earlier {[}see also the cartoon in Fig.\ref{fig:Pic1}(g){]}.
This feature strongly suggests that the atoms from the dimer does
not constitute a molecule anymore and can be considered, in principle,
decoupled from each other. As a matter of fact, in Fig.\ref{fig:Pic3}
we will see that such a single peak is not atomic as it seems to be,
as pointed out by the on-site spectral function $\rho_{jj}^{\sigma}(\omega)$
pattern. Surprisingly, such a case is peculiar and it has a molecular
nature, as can be ensured by the finite values extracted from the
molecular correlation function $\delta\rho_{jl}(\omega)$ ($j\neq l=1,2$)
in Eq.(\ref{eq:LDOSjl}). This nonlocal quantity then reveals that
the molecular binding persists and due to such, we recognize this
molecular zero mode as atomically frustrated.

In Fig.\ref{fig:Pic3}(a) for the first twist magic angle condition
$\theta=\theta_{M}=1.08^{{^{\circ}}},$ we summarize via the $\rho_{jj}^{\sigma}(\omega)$
color map (in normalized units) as a function of $V$ and $\omega,$
the crossover from ordinary molecular states towards a molecular zero
mode atomically frustrated. This density plot makes explicit three
distinct phases, which we recognize as: I) the weak metallic Moiré
phase, where four molecular states can be viewed corresponding to
the splittings in the Hubbard bands $\varepsilon_{dj\sigma}<0$ and
$\varepsilon_{dj\sigma}+U_{j}>0,$ II) the intermediate metallic phase
with two atomic frustrated states well established as in Weyl
metals\cite{Frustration} and finally, III) the strong metallic Moiré
phase, where the molecular zero mode atomically frustrated emerges.
{Although we have showed in Fig.\ref{fig:Pic2}(f) that the gate $V=0.03D$ between
the layers affects the energy spectrum of the atomic impurities with the presence of the zero mode when $\theta=\theta_{M}=1.08^{{^{\circ}}},$  in the $\theta=1.07^{{^{\circ}}}$ case, this single peak does not exist for such a $V$ strength. However, for $\theta=\theta_{M}=1.08^{{^{\circ}}},$ the zero mode threshold to rise is much lower, i.e., it is given by $V=0.019D,$ as marked at Fig.\ref{fig:Pic3}(a) in rose color. Such a minor value is expected, since our findings lie within the domains of linear dispersion approximation and nearly flat bands.}
Thus, to better understand the reason for which the absence of splittings
in the atomic levels observed in $\rho_{jj}^{\sigma}(\omega)$ is
considered naive to characterize the dimer's atoms dissociated and
the final state as a genuine atomic one, we call attention to the molecular
correlation function $\delta\rho_{jl}(\omega)$ ($j\neq l=1,2$) as
a function of $\omega.$ Notice that in Fig.\ref{fig:Pic3}(b) for
$V=0,$ which corresponds to the split bonding and antibonding molecular
levels in Fig.\ref{fig:Pic2}(d), the $\delta\rho_{jl}(\omega)$ spectral
function is finite (red lineshape), as expected for a molecular binding.
The same occurs for $V=0.01D$ (green lineshape) and $V=0.03D$ (blue
lineshape). However, the latter leads to only one single resonant structure
at $\omega=0$ (zero mode) in Fig.\ref{fig:Pic2}(f), which could
give the impression that the atoms become decoupled from each other,
turning into authentic atomic states. As $\delta\rho_{jl}(\omega)$
is finite even with $V=0.03D,$ the Moiré superlattice intermediates
the connection between the atoms, just in order to sustain the molecular
binding. Particularly, this mediation is assisted by the atomic Coulomb
interactions $U_{j},$ as we can verify in the term $\lambda_{j}^{\bar{\sigma}}{\Sigma_{0}(\omega)g_{j\sigma|l\sigma}(\omega)\lambda_{l}^{\bar{\sigma}}\Sigma_{0}(\omega)}$
($j\neq l=1,2$) for Eq.(\ref{eq:Gjj}). Its origin arises from the
interplay between the fulfilment of the first twist magic angle condition
$\theta=\theta_{M}=1.08^{{^{\circ}}}$ in the presence of Coulomb
terms and the breaking down of the inversion symmetry in the Dirac cones by the gate
voltage $V.$ To conclude, we analyze in Fig.\ref{fig:Pic3}(c) how the \textit{quasiparticle}
lifetime $\tau(U_{j})$\cite{ManyBody} for the molecular zero mode
in the atomic frustrated state regime $(V=0.03D)$ changes with the
Coulomb interactions $U_{j}.$ Thus, we restore in the profile of $\rho_{jj}^{\sigma}(\omega)$
the positive energy domain $(\omega>0)$, once the lifetime is inversely
proportional to the spectral width $2\Gamma(U_{j})$\cite{ManyBody},
i.e., $\tau(U_{j})=\frac{\hbar}{2\Gamma(U_{j})}.$ Parameterizing
by $\varepsilon_{dj\sigma}=-0.01fD$ and $U_{j}=-2\varepsilon_{dj\sigma}$,
we find out that the \textit{quasiparticle} lifetime increases by
decreasing $f.$ It means that, hypothetically in the absence of such
an interaction ($f=0$), the dimer's atoms become isolated from the
superlattice and the atomic frustrated state disappears, due to the
$\rho_{jj}^{\sigma}(\omega)\approx\delta(\omega)$ behavior {[}panel
(c)-I of the same figure{]}. Therefore, the molecular binding found
in this work should be afforded by a finite $U_{j}$ to exist. Interestingly
enough, the Anderson theory on impurity systems\cite{Anderson1961}
predicts spectral densities with single peaks solely in the absence
of Coulomb interactions. Such a signature, as we can notice, arises
rightly from the Hubbard-I solution, which by imposing $U_{j}=0$
in Eq.(\ref{eq:HB_I}), provides a spectral density with a \textit{Lorentzian}
lineshape. However, the molecular zero mode of the atomic frustrated
state here reported, which is also a single one resonant peak of a
\textit{Lorentzian}, then contrasts with the aforementioned feature
and constitutes an exception to this rule. As a matter of fact, the
results from Fig.\ref{fig:Pic3}(c) ensure that such a state is indeed,
a strongly correlated metallic state, which emerges at the first twist
magic angle condition.

\section{Conclusions}

We have demonstrated that the TBG is a promising candidate to host
a molecular zero mode atomically frustrated, upon considering diatomic
molecules with charge mirror symmetry. To realize such, in particular,
at the first twist magic angle, the TBG system should be driven into
a strong metallic phase, which arises from the broken inversion symmetry
in the Dirac cones for the emergent mini Brillouin zones of the Moiré superlattice.
This can be performed by applying top and bottom opposite gate voltages
to the graphene monolayers. By this manner, this configuration leads
to an enhancement of the blue and red shifts in the molecular levels,
which in the case of considering atomic Coulomb correlations, make
the possibility of merging entirely these molecular peaks into a zero
mode in the dimer's atoms, just by changing the gate voltages. These
tunable states constitute a spectral single peak, which suggests that the atoms
are found decoupled from each other. However, the molecular correlation function from the dimer's atoms
is still finite in this case, thus revealing the molecular nature
of such a binding, which in particular, is mediated by the Moiré superlattice. As a result,
this leads then to the concept of atomic frustration in molecules.
In summary, our findings point out that the Moiré superlattice can
be considered as a new platform to build unconventional molecular
bindings via graphene twistronics.

\section{Acknowledgments}

We thank the Brazilian funding agencies CNPq (Grants. Nos. 128919/2021-3, 308410/2018-1, 305668/2018-8
and 302887/2020-2), the São Paulo Research Foundation (FAPESP; Grant No. 2018/09413-0) and Coordenação de Aperfeiçoamento de Pessoal de
Nível Superior - Brasil (CAPES) -- Finance Code 001.

\appendix

\section*{Appendices}

We summarize the main mathematical steps to derive the GFs of the
dimer's atoms {[}Eqs.(\ref{eq:Gjj}) and (\ref{eq:Gjl}){]} and the
molecular correlation function {[}Eq.(\ref{eq:LDOSjl}){]}. Such evaluations
will be performed via the equation-of-motion (EOM) approach on the
system GFs\cite{ManyBody}, in particular, within the Hubbard-I approximation\cite{Hubbard1963}.

\section{Helical basis for the TBG Anderson-like Hamiltonian }

In order to accomplish the previous announced purpose, we first need
to diagonalize the Hamiltonian of the TBG system and later on, express
it into the helical basis with right- and left-movers. Thus, by using
the spinor $\psi_{\text{\textbf{k}}\xi\sigma}^{\dagger}=(\begin{array}{cc}
\psi_{AB\text{\textbf{k}}\xi\sigma}^{\dagger} & \psi_{BA\text{\textbf{k}}\xi\sigma}^{\dagger}\end{array})$ in Eq.(\ref{eq:Hamiltonian}), we begin with
\begin{eqnarray}
\mathcal{H} & = & \mathcal{H}_{\text{{TBG}}}+\mathcal{H}_{\text{{dimer}}}+\mathcal{H}_{\text{{Hyb}}}\nonumber \\
 & = & \sum_{\text{\textbf{k}}\xi\sigma}[\hbar\tilde{v}_{F}(\xi k_{x}-ik_{y})+\xi V]\psi_{AB\text{\textbf{k}}\xi\sigma}^{\dagger}\psi_{BA\text{\textbf{k}}\xi\sigma}+\text{H.c.}\nonumber \\
 & + & \sum_{j\sigma}\varepsilon_{dj\sigma}d_{j\sigma}^{\dagger}d_{j\sigma}+\sum_{j}U_{j}d_{j\uparrow}^{\dagger}d_{j\uparrow}d_{j\downarrow}^{\dagger}d_{j\downarrow}\nonumber \\
 & + & \sum_{\text{\textbf{k}}\xi j\sigma}v_{0}(\psi_{AB\text{\textbf{k}}\xi\sigma}^{\dagger}d_{j\sigma}+\text{H.c.}).\nonumber \\
\label{eq:FullHamiltonian}
\end{eqnarray}
Now we change to polar coordinates $\hbar\tilde{v}_{F}(\xi k_{x}\mp ik_{y})=\hbar\tilde{v}_{F}k\xi e^{\mp i\xi\theta},$
perform the transformations $\psi_{AB\text{\textbf{k}}\xi\sigma}=\frac{1}{\sqrt{2}}[\zeta_{+\sigma}^{\xi}(\text{\textbf{k}})+\zeta_{-\sigma}^{\xi}(\text{\textbf{k}})],$ $\psi_{BA\text{\textbf{k}}\xi\sigma}=\frac{\xi}{\sqrt{2}}e^{i\xi\theta}[\zeta_{+\sigma}^{\xi}(\text{\textbf{k}})-\zeta_{-\sigma}^{\xi}(\text{\textbf{k}})]$
and $\sum_{\text{\textbf{k}}}\rightarrow\frac{\Omega_{\theta}}{\left(2\pi\right)^{2}}\int d^{2}\text{\textbf{k}}$ in the Hamiltonians $\mathcal{H}_{\text{{TBG}}}$ and $\mathcal{H}_{\text{{Hyb}}},$
just to obtain the continuum limit of \textbf{k} space. Thus,
by taking into account all these procedures together, we find
\begin{eqnarray}
\mathcal{H}_{\text{{TBG}}} & = & \sum_{\xi\sigma}\frac{\Omega_{\theta}}{2\pi}[\int_{0}^{\infty}\varepsilon_{k}kdk(\zeta_{+\sigma}^{\xi\dagger}(\text{\textbf{k}})\zeta_{+\sigma}^{\xi}(\text{\textbf{k}})\nonumber \\
 & - & \zeta_{-\sigma}^{\xi\dagger}(\text{\textbf{k}})\zeta_{-\sigma}^{\xi}(\text{\textbf{k}}))+\xi V(\zeta_{+\sigma}^{\xi\dagger}(\text{\textbf{k}})\zeta_{+\sigma}^{\xi}(\text{\textbf{k}})\nonumber \\
 & + & \zeta_{-\sigma}^{\xi\dagger}(\text{\textbf{k}})\zeta_{-\sigma}^{\xi}(\text{\textbf{k}}))],\label{eq:Hamiltonian_1}
\end{eqnarray}
with $\varepsilon_{k}=\hbar\tilde{v}_{F}k$ and
\begin{eqnarray}
\mathcal{H}_{\text{{Hyb}}} & = & \sum_{\xi j\sigma}\frac{\Omega_{\theta}}{2\pi}\int_{0}^{\infty}kdk\frac{v_{0}}{\sqrt{2}}(\zeta_{+\sigma}^{\xi\dagger}(\text{\textbf{k}})+\zeta_{-\sigma}^{\xi\dagger}(\text{\textbf{k}}))d_{j\sigma}+\text{H.c..}\nonumber \\
\label{eq:Hamiltonian_2}
\end{eqnarray}
As we are interested in the low energy Physics of the TBG system,
which should be isotropic as expected, we perform an angular plane-wave
decomposition of the form $\zeta_{\pm\sigma}^{\xi}(\text{\textbf{k}})=\frac{1}{\sqrt{k}}[\frac{(2\pi)^{2}}{\Omega_{\theta}}]^{\frac{1}{2}}\sum_{m=-\infty}^{\infty}\frac{1}{\sqrt{2\pi}}$
$e^{im\theta}\zeta_{\pm\sigma\xi}^{m}(k),$ from where we only catch
the contribution of the $m=0$ state. Later on, we unfold the range
of the momenta $k$ from $(0,\infty)$ to $(-\infty,+\infty),$ which
allows us to introduce the helical basis $c_{\sigma\xi}(k)=\zeta_{+\sigma}^{\xi}(k)$
for $k>0$ and $c_{\sigma\xi}(k)=\zeta_{-\sigma}^{\xi}(k)$ for $k<0$
to describe right- and left-movers, respectively. Consequently, we
find an Anderson-like Hamiltonian\cite{Anderson1961} for the TBG
system, which reads
\begin{eqnarray}
\mathcal{H} & = & \sum_{\xi\sigma}\int_{-\infty}^{+\infty}dk(\varepsilon_{k}+\xi V)c_{\sigma\xi}^{\dagger}(k)c_{\sigma\xi}(k)+\mathcal{H}_{\text{{dimer}}}\nonumber \\
 & + & \sum_{\xi j\sigma}\int_{-\infty}^{+\infty}dk\mathcal{V}_{k}(\theta)(c_{\sigma\xi}^{\dagger}(k)d_{j\sigma}+\text{H.c.}),\label{eq:Complete}
\end{eqnarray}
where $\mathcal{V}_{k}(\theta)=\frac{v_{0}\sqrt{\Omega_{\theta}|k|\pi}}{2\pi}$
encodes the momentum and twist angle dependent coupling between the
impurity atoms and the Moiré superlattice.

\section{The evaluation of the GFs for the dimer's atoms}

The EOM approach can be summarized as follows: $\omega^{+}\tilde{{\cal G}}_{\mathcal{A}|\mathcal{B}}(\omega)=\left\{ \mathcal{A},\mathcal{B}\right\} +\tilde{{\cal G}}_{[\mathcal{A},\mathcal{H}]|\mathcal{B}}(\omega),$
$\omega^{+}=\omega+i0^{+},$ the time Fourier transform $\tilde{{\cal G}}_{\mathcal{A}|\mathcal{B}}(\omega)=\int_{-\infty}^{+\infty}dt\mathcal{G}_{\mathcal{A}|\mathcal{B}}(t)e^{-\frac{i}{\hbar}\omega^{+}t}$
for the energy domain GF $\tilde{{\cal G}}_{\mathcal{A}|\mathcal{B}}(\omega)$
and $\mathcal{G}_{\mathcal{A}|\mathcal{B}}(t)=-\frac{i}{\hbar}\theta(t)<\{\mathcal{A}(t),\mathcal{B}{}^{\dagger}(0)\}>_{\mathcal{H}}$
is the corresponding GF in time domain. To obtain Eqs.(\ref{eq:Gjj})
and (\ref{eq:Gjl}), we follow the EOM prescription taking into account Eq.(\ref{eq:Complete}) and show
\begin{eqnarray}
 &  & (\omega^{+}-\varepsilon_{dj\sigma})\tilde{{\cal G}}_{{j\sigma}|{{l}\sigma}}(\omega)=\delta_{jl}+U_{j}\tilde{{\cal G}}_{{j\sigma}n_{j\bar{\sigma}}|{{l}\sigma}}(\omega)\nonumber \\
 &  & +\sum_{\xi}\int_{-\infty}^{+\infty}dk\mathcal{V}_{k}(\theta)\tilde{{\cal G}}_{c_{\sigma\xi}(k)|{{l}\sigma}}(\omega),\label{eq:GF1}
\end{eqnarray}
with $\mathcal{G}_{{j\sigma}n_{j\bar{\sigma}}|{{l}\sigma}}(t)=-\frac{i}{\hbar}\theta(t)<\{d_{j\sigma}(t)n_{j\bar{\sigma}}(t),d_{l\sigma}^{\dagger}(0)\}>_{\mathcal{H}}$
as a GF of high hierarchy that encodes the on-site Coulomb correlations,
with $n_{j\bar{\sigma}}=d_{j\bar{\sigma}}^{\dagger}d_{j\bar{\sigma}},$
while the host-atom mixing GF arises from
\begin{eqnarray}
\tilde{{\cal G}}_{c_{\sigma\xi}(k)|{{l}\sigma}}(\omega) & = & \sum_{j'}\frac{\mathcal{V}_{k}(\theta)}{(\omega^{+}-\varepsilon_{k}-\xi V)}\tilde{{\cal G}}_{{j'\sigma}|{{l}\sigma}}(\omega),\label{eq:GF2}
\end{eqnarray}
in which $\mathcal{G}_{c_{\sigma\xi}(k)|{{l}\sigma}}(t)=-\frac{i}{\hbar}\theta(t)<\{c_{\sigma\xi}(k,t),d_{l\sigma}^{\dagger}(0)\}>_{\mathcal{H}}.$
From Eq.(\ref{eq:GF1}), we need to obtain $\tilde{{\cal G}}_{{j\sigma}n_{j\bar{\sigma}}|{{l}\sigma}}(\omega).$
As we are not interested in the Kondo regime\cite{Kondo}, spin-flip
processes can be safely disregarded in the evaluation of $\tilde{{\cal G}}_{{j\sigma}n_{j\bar{\sigma}}|{{l}\sigma}}(\omega)$.
Such an assumption consists the main cornerstone of the Hubbard-I
approach in decoupling the system of GFs\cite{Hubbard1963}. The Hubbard-I
holds at temperatures $T\gg T_{K}$, where $T_{K}$ represents the
Kondo temperature\cite{Kondo}. Particularly in the calculation of
$\bigl\langle n_{j\bar{\sigma}}\bigr\rangle$ given by Eq.(\ref{eq:Number})
and below, we make explicit that $T$ should not be very high, so
that we can safely adopt the Heaviside step function for the Fermi-Dirac
distribution $n_{_{F}}(\varepsilon).$ Thus, by this manner, after
successive steps of the EOM, we obtain
\begin{eqnarray}
 &  & (\omega^{+}-\varepsilon_{dj\sigma}-U_{j})\tilde{{\cal G}}_{{j\sigma}n_{j\bar{\sigma}}|{{l}\sigma}}(\omega)=\delta_{jl}<n_{j\bar{\sigma}}>\nonumber \\
 &  & +\sum_{\xi}\int_{-\infty}^{+\infty}dk\mathcal{V}_{k}(\theta)\tilde{{\cal G}}_{c_{\sigma\xi}(k)n_{j\bar{\sigma}}|{{l}\sigma}}(\omega),\label{eq:GF3}
\end{eqnarray}
$\mathcal{G}_{c_{\sigma\xi}(k)n_{j\bar{\sigma}}|{{l}\sigma}}(t)=-\frac{i}{\hbar}\theta(t)<\{c_{\sigma\xi}(k,t)n_{j\bar{\sigma}}(t),d_{l\sigma}^{\dagger}(0)\}>_{\mathcal{H}}$
and
\begin{eqnarray}
 &  & (\omega^{+}-\varepsilon_{k}-\xi V)\tilde{{\cal G}}_{c_{\sigma\xi}(k)n_{j\bar{\sigma}}|{{l}\sigma}}(\omega)=\sum_{j'}\mathcal{V}_{k}(\theta)\tilde{{\cal G}}_{{j'\sigma}n_{j\bar{\sigma}}|{{l}\sigma}}(\omega),\nonumber \\
\label{eq:GF4}
\end{eqnarray}
with $\mathcal{G}_{{j'\sigma}n_{j\bar{\sigma}}|{{l}\sigma}}(t)=-\frac{i}{\hbar}\theta(t)<\{d_{j'\sigma}(t)n_{j\bar{\sigma}}(t),d_{l\sigma}^{\dagger}(0)\}>_{\mathcal{H}}.$
At this stage, we impose the mean-field approximation $\tilde{{\cal G}}_{{j'\sigma}n_{j\bar{\sigma}}|{{l}\sigma}}(\omega)\approx<n_{j\bar{\sigma}}>\tilde{{\cal G}}_{{j'\sigma}|{{l}\sigma}}(\omega),$
which allows us to close the set of equations for the system of GFs.
This gives rise to Eqs.(\ref{eq:Gjj}) and (\ref{eq:Gjl}), wherein
\begin{equation}
\Sigma_{0}^{\xi}(\omega)=\int_{-D}^{+D}\frac{d\varepsilon_{k}}{\hbar\tilde{v}_{F}}\frac{\mathcal{V}_{k}^{2}(\theta)}{(\omega^{+}-\varepsilon_{k}-\xi V)}\label{eq:SE1}
\end{equation}
determines the self-energy in Eq.(\ref{eq:SE0}), with $\pm D$ as
the ultraviolet $(+)$ and infrared energy-cutoffs $(-)$, respectively.

\section{The molecular correlation function}

As previously stated in the main text, the correlation function $\delta\rho_{jl}(\omega)$
in $\sum_{jl}\delta\rho_{jl}(\omega)=-\frac{1}{\pi}\text{{Im}}[\tilde{\mathcal{G}}^{\text{{corr}}}(\omega)]$
($j\neq l=1,2$) requires the determination of the GF $\tilde{\mathcal{G}}^{\text{{corr}}}(\omega),$
which is obtained from the time Fourier transform of $\mathcal{G}^{\text{{corr}}}(t)=-\frac{i}{\hbar}\theta(t)<\{f_{0\sigma}(t),f_{0\sigma}^{\dagger}(0)\}>_{\mathcal{H}}.$
Thus, the operator $f_{0\sigma}$ describes a host site from which the molecular
binding can be established by connecting the atoms through the TBG
system. Particularly, in the helical basis $f_{0\sigma}=\sum_{\xi}\int_{-\infty}^{+\infty}dk\frac{\sqrt{\Omega_{\theta}|k|\pi}}{2\pi}c_{\sigma\xi}(k)$ according to Eq.(\ref{eq:Complete}) and
\begin{eqnarray}
 & \sum_{jl}\delta\rho_{jl}(\omega) & =\sum_{\xi\xi'\sigma}(\frac{\sqrt{\Omega_{\theta}\pi}}{2\pi})^{2}\int_{-\infty}^{+\infty}\int_{-\infty}^{+\infty}dkdq\sqrt{|k|}\sqrt{|q|}\nonumber \\
 &  & \times(-\frac{1}{\pi})\text{{Im}[}\tilde{{\cal G}}_{{c_{\sigma\xi}(k)}|{c_{\sigma\xi'}(q)}}(\omega)]\label{eq:rhojl}
\end{eqnarray}
expressed by the last GF, which in time domain reads $\mathcal{G}_{{c_{\sigma\xi}(k)}|{c_{\sigma\xi'}(q)}}(t)=-\frac{i}{\hbar}\theta(t)<\{c_{\sigma\xi}(k,t),c_{\sigma\xi'}^{\dagger}(q,0)\}>_{\mathcal{H}}.$
By applying the EOM to unknown GFs, we find
\begin{eqnarray}
 &  & \tilde{{\cal G}}_{{c_{\sigma\xi}(k)}|{c_{\sigma\xi'}(q)}}(\omega)=\frac{\delta_{\xi\xi'}\delta(k-q)}{(\omega^{+}-\varepsilon_{k}-\xi V)}\nonumber \\
 &  & +\sum_{j}\frac{v_{0}\sqrt{\Omega_{\theta}|k|\pi}}{2\pi(\omega^{+}-\varepsilon_{k}-\xi V)}\tilde{{\cal G}}_{{j\sigma}|c_{\sigma\xi'}(q)}(\omega),\label{eq:GF5}
\end{eqnarray}
where $\mathcal{G}_{{j\sigma}|c_{\sigma\xi'}(q)}(t)=-\frac{i}{\hbar}\theta(t)<\{d_{j\sigma}(t),c_{\sigma\xi'}^{\dagger}(q,0)\}>_{\mathcal{H}}$
and
\begin{eqnarray}
\tilde{{\cal G}}_{{j\sigma}|c_{\sigma\xi'}(q)}(\omega) & = & \sum_{l}\frac{v_{0}\sqrt{\Omega_{\theta}|q|\pi}}{2\pi(\omega^{+}-\varepsilon_{q}-\xi'V)}\tilde{{\cal G}}_{{j\sigma}|{{l}\sigma}}(\omega).\nonumber \\
\label{eq:GF6}
\end{eqnarray}
Therefore, after substituting Eqs.(\ref{eq:GF5}) and (\ref{eq:GF6})
into Eq.(\ref{eq:rhojl}), we finally deduce the molecular correlation
function given by Eq.(\ref{eq:LDOSjl}), which should consider $j\neq l=1,2$.


\begin{thebibliography}{1}
\bibitem{Herrero1}{Y. Cao, V. Fatemi, S. Fang, K. Watanabe, T. Taniguchi,
E. Kaxiras, and P. Jarillo-Herrero, Unconventional Superconductivity
in Magic-Angle Graphene Superlattices, Nature (London) 556, 43 (2018).}

\bibitem{Herrero2}{Y. Cao, V. Fatemi, A. Demir, S. Fang, S. L. Tomarken,
J. Y. Luo, J. D. Sanchez-Yamagishi, K. Watanabe, T. Taniguchi, E.
Kaxiras, R. C. Ashoori, and P. Jarillo-Herrero, Correlated Insulator
Behaviour at Half-Filling in Magic-Angle Graphene Superlattices, Nature
(London) 556, 80 (2018).}

\bibitem{T1}{M. Fidrysiak, M. Zegrodnik, and J. Spalek, Unconventional
topological superconductivity and phase diagram for an effective two-orbital
model as applied to twisted bilayer graphene, Phys. Rev. B 98, 085436
(2018).}

\bibitem{T2}{T. J. Peltonen, R. Ojajärvi, and T. T. Heikkilä, Mean-field
theory for superconductivity in twisted bilayer graphene, Phys. Rev.
B 98, 220504 (2018).}

\bibitem{T3}{X. Y. Xu, K. T. Law, and P. A. Lee, Kekulé valence
bond order in an extended Hubbard model on the honeycomb lattice with
possible applications to twisted bilayer graphene, Phys. Rev.B 98,
121406 (2018).}

\bibitem{T4}{C.-C. Liu, L.-D. Zhang, W.-Q. Chen, and F. Yang, Chiral
Spin Density Wave and d+id Superconductivity in the Magic-Angle-Twisted
Bilayer Graphene, Phys. Rev. Lett. 121, 217001 (2018).}

\bibitem{T5}{S. Ray and T. Das, Wannier pairs in superconducting
twisted bilayer graphene and related systems, Phys. Rev. B 99, 134515
(2019).}

\bibitem{T6}{J. F. Dodaro, S. A. Kivelson, Y. Schattner, X.-Q. Sun,
and C. Wang, Phases of a phenomenological model of twisted bilayer
graphene, Phys. Rev. B 98, 075154 (2018).}

\bibitem{T7}{B. Padhi, C. Setty, and P. W. Phillips, Doped Twisted
Bilayer Graphene near Magic Angles: Proximity to Wigner Crystallization
not Mott Insulation, Nano Lett. 2018 18 (10) 6175.}

\bibitem{T8}{H. Guo, X. Zhu, S. Feng, and R. T. Scalettar, Pairing
Symmetry of Interacting Fermions on Twisted Bilayer Graphene Superlattice,
Phys. Rev. B 97, 235453 (2018).}

\bibitem{T9}{B. Roy and V. Juricic, Unconventional superconductivity
in nearly flat bands in twisted bilayer graphene, Phys. Rev. B 99,
121407 (2019).}

\bibitem{T10}{H. C. Po, L. Zou, A. Vishwanath, and T. Senthil, Origin
of Mott Insulating Behavior and Superconductivity in Twisted Bilayer
Graphene, Phys. Rev. X 8, 031089 (2018).}

\bibitem{T11}{N. F. Q. Yuan and L. Fu, A Model for Metal-Insulator
Transition in Graphene Superlattices and Beyond, Phys. Rev. B 98,
045103 (2018).}

\bibitem{T12}{C. Xu and L. Balents, Topological Superconductivity
in Twisted Multilayer Graphene, Phys. Rev. Lett. 121, 087001 (2018).}

\bibitem{EXP1}{A. Luican, G. Li, A. Reina, J. Kong, R. R. Nair,
K. S. Novoselov, A. K. Geim, and E. Y. Andrei, Single-Layer Behavior
and Its Breakdown in Twisted Graphene Layers, Phys. Rev. Lett. 106,
126802 (2011).}

\bibitem{EXP2}{D. L. Miller, K. D. Kubista, G. M. Rutter, M. Ruan,
W. A. de Heer, P. N. First, and J. A. Stroscio, Structural Analysis
of Multilayer Graphene via Atomic Moiré Interferometry, Phys. Rev.
B 81, 125427 (2010).}

\bibitem{EXP3}{G. Li, A. Luican, J. M. B. Lopes dos Santos, A. H.
C. Neto, A. Reina, J. Kong, and E. Y. Andrei, Observation of van Hove
Singularities in Twisted Graphene Layers, Nat. Phys. 6, 109 (2010).}

\bibitem{TT1}{P. Moon and M. Koshino, Optical Absorption in Twisted
Bilayer Graphene, Phys. Rev. B 87, 205404 (2013).}

\bibitem{TT2}{G. Trambly de Laissardiere, D. Mayou, and L. Magaud,
Numerical Studies of Confined States in Rotated Bilayers of Graphene,
Phys. Rev. B 86, 125413 (2012).}

\bibitem{TT3}{P. Moon and M. Koshino, Energy Spectrum and Quantum
Hall Effect in Twisted Bilayer Graphene, Phys. Rev. B 85, 195458 (2012).}

\bibitem{CastroNetoPRB}{J. M. B. Lopes dos Santos, N. M. R. Peres,
and A. H. Castro Neto, Continuum Model of the Twisted Graphene Bilayer,
Phys. Rev. B 86, 155449 (2012).}

\bibitem{TT5}{L. Xian, S. Barraza-Lopez, and M. Y. Chou, Effects
of Electrostatic Fields and Charge Doping on the Linear Bands in Twisted
Graphene Bilayers, Phys. Rev. B 84, 075425 (2011).}

\bibitem{TT6}{M. Kindermann and P. N. First, Local Sublattice-Symmetry
Breaking in Rotationally Faulted Multilayer Graphene, Phys. Rev. B
83, 045425 (2011).}

\bibitem{TT7}{R. Bistritzer and A. H. MacDonald, Moiré Bands in
Twisted Double-Layer Graphene, Proc. Natl. Acad. Sci. U.S.A. 108,
12233 (2011).}

\bibitem{TT8}{E. S. Morell, J. D. Correa, P. Vargas, M. Pacheco,
and Z. Barticevic, Flat Bands in Slightly Twisted Bilayer Graphene:
Tight-Binding Calculations, Phys. Rev. B 82, 121407 (2010).}

\bibitem{TT9}{G. Trambly de Laissardière, D. Mayou, and L. Magaud,
Localization of Dirac Electrons in Rotated Graphene Bilayers, Nano
Lett. 10, 804 (2010).}

\bibitem{TT10}{E. J. Mele, Commensuration and Interlayer Coherence
in Twisted Bilayer Graphene, Phys. Rev. B 81, 161405 (2010).}

\bibitem{Review}{A. V. Rozhkov, A. O. Sboychakov, A. L. Rakhmanov,
and F. Nori, Electronic properties of graphene-based bilayer systems,
Physics Reports, 648, 1-104 (2016).}

\bibitem{CastroNetoPRL}{J. M. B. Lopes dos Santos, N. M. R. Peres,
and A. H. Castro Neto, Graphene Bilayer with a Twist: Electronic Structure,
Phys. Rev. Lett. 99, 256802 (2007).}

\bibitem{EJMele}{Võ Tiên Phong and E. J. Mele, Obstruction and Interference
in Low Energy Models for Twisted Bilayer Graphene, Phys. Rev. Lett.
125, 176404 (2020).}

\bibitem{Frustration}{W. N. Mizobata, Y. Marques, M. Penha, J. E.
Sanches, L. S. Ricco, M. de Souza, I. A. Shelykh, and A. C. Seridonio,
Atomic frustrated impurity states in Weyl metals, Phys. Rev. B 102,
075120 (2020).}

\bibitem{Dirac}{Y.Marques, A.E. Obispo, L.S. Ricco, M. de Souza,
I.A. Shelykh, and A.C. Seridonio, Antibonding Ground state of Adatom
Molecules in Bulk Dirac Semimetals, Phys. Rev. B 96, 041112 (2017).}

\bibitem{Weyl}{Y. Marques, W. N. Mizobata, R. S. Oliveira, M. de
Souza, M. S. Figueira, I. A. Shelykh, and A. C. Seridonio, Chiral
magnetic chemical bonds in molecular states of impurities in Weyl
semimetals, Scientific Reports 9, 8452 (2019).}

\bibitem{Majorana1}{K. Flensberg, F. von Oppen, and A. Stern, Engineered
platforms for topological superconductivity and Majorana zero modes, Nat. Rev. Mater.
https://doi.org/10.1038/s41578-021-00336-6 (2021).}

\bibitem{Majorana2}{B. Jäck, Y. Xie, and A. Yazdani, Detecting and distinguishing Majorana
zero modes with the scanning tunnelling microscope, Nat. Rev. Phys. 3, 541 (2021).}

\bibitem{Anderson1961}{P. W. Anderson, Localized Magnetic States
in Metals, Phys. Rev. 124, 41 (1961).}

\bibitem{Cluster}{Y. Wang, D. Wong, A. V. Shytov, V. W. Brar, S. Choi, Q. Wu,
H.-Z. Tsai, W. Regan, A. Zettl, R. K. Kawakami, S. G. Louie, L. S. Levitov, and M. F. Crommie, Observing Atomic Collapse
Resonances in Artificial Nuclei on Graphene, Science 340, 734 (2013).}

\bibitem{STM1}{I. Brihuega, P. Mallet, H. González-Herrero, G. Trambly
de Laissardière, M.M. Ugeda, L. Magaud, J.M. Gómez-Rodríguez, F. Ynduráin,
J.-Y. Veuillen, Unraveling the intrinsic and robust nature of van
Hove singularities in twisted bilayer graphene by scanning tunneling
microscopy and theoretical analysis, Phys. Rev. Lett. 109, 196802
(2012).}

\bibitem{STM2}{E. Cisternas, J. Correa, Theoretical reproduction
of superstructures revealed by STM on bilayer graphene, Chem. Phys.
409, 74 (2012).}

\bibitem{Hubbard1963}{J. Hubbard, Electron Correlations in Narrow
Energy Bands, Proc. R. Soc. A 276, 238 (1963).}

\bibitem{Kondo}{A. C. Hewson, The Kondo problem to Heavy Fermions,
Cambridge University Press, (1993).}

\bibitem{Puddles}{S. Zhang, D. Huang, L. Gu, Y. Wang, and S. Wu, Substrate dopant induced electronic inhomogeneity in epitaxial
bilayer graphene, 2D Mater. 8, 035001 (2021).}

\bibitem{Doping}{H. Lee, K. Paeng, and I. S. Kim, A review of doping modulation in graphene, Synthetic Metals 244, 36-47 (2018).}

\bibitem{ManyBody}{H. Bruus and K. Flensberg, Many-Body Quantum
Theory in Condensed Matter Physics, An Introduction, Oxford University
Press, (2012).}

\end{thebibliography}
\end{document}